# INTERNATIONAL CO-BRANDING
# AND FIRM'S FINANCIAL PERFORMANCE

**Hengameh Fakhravar**
**Engineering Management & Systems Engineering**
**Old Dominion University**

**Hesamoddin Tahami**
**Engineering Management & Systems Engineering**
**Old Dominion University**

**\*Hfakhrav@odu.edu**

___________________________________________________________________

**Abstract**

Co-branding has become a widely used marketing strategy, yet little attention has been paid to its impact on a firm's stock value. Prior literature has shown that using a co-branding strategy properly helps firms leverage brand value and equity. We discussed the theoretical foundations and main accomplishments of prior studies. We developed a conceptual framework and hypothesis to close the existing research gap in the topic of interest. We argued that co-branding event announcement generates positive abnormal returns in the stock market. Furthermore, we investigated the moderating impact of co-branding structure on the relation between co-branding event announcements and abnormal returns. We claimed that higher co-branding integration, greater co-branding exclusivity, and longer co-branding duration generate a greater positive abnormal return for the partnering firms.

## 1- Introduction

Brands are the most valuable assets of companies pursuing commercial success ((Keränen et al., 2012); (Paydas Turan, 2021)). As markets become increasingly competitive and consumers ever more demanding, business managers seek out ways of enhancing customer experience and business growth and are therefore turning to co-branding (Besharat & Langan, 2014). Co-branding represents a long-term brand alliance strategy in which one product is branded and identified simultaneously by two brands. From Dell Computers with Intel Processors to Kellogg Star Wars cereal to Philips shavers dispensing Nivea shaving cream, co-branded products take many forms across industries, at times connecting seemingly unlikely alliance partners.

Co-branding combines the individual brand characteristics of the constituent brands and transfers the associated values of both brands to the created co-branding (Yu et al., 2021)(Yahoodik et al., 2020). The purpose of forming a co-branding is to create synergies, boost awareness, and enhance the brands' value by leveraging each brand through the transfer of associations from one brand to the other and differentiating them from the competition (Singh et al., 2016). According to this definition, the following characteristics constitute co-branded products: First, the participating brands should be independent before, during, and after the offering of the co-branded product (Ohlwein & Schiele, 1994). Second, the companies that own the brands should implement a co-branding strategy on purpose (Blackett & Russell, 1999).

Third, the cooperation between the two brands must be visible to potential buyers (Rao, 1997), and fourth, one product must be combined with the two other brands at the same time (Levin et al., 1996)(Fakhravar, 2020b).

However, despite practitioners' apparent enthusiasm towards co-branded products, research has not yet determined if these are profitable investments for their parent firms. Indeed, co-branding may also have downsides. Co-branding carries the risk of eroding brand equity through potentially inconsistent brand associations and potential loss in perceived quality. Negative associations can transfer from partner brands to the co-branded product, hindering its market success. Alternatively, negative associations can also transfer from the co-branded product to one of the partner brands. For instance, in an experiment intended to assess preferences for brownies made from a co-branded mix, (Levin et al., 1996) found that if one partner brand is thought to be inferior (in their case, the brand of chocolate chips used in the brownie mix), it brings down not only the perception of the co-branded product but also that of the other partner brand. In addition to risk, there might also be direct costs that are greater with co-branding, for example, coordination efforts between partner firms.

This study examines the financial consequences of co-branding activities and illuminates some key questions left unexplored by previous research. First, while co-branding can improve customers' attitudes towards the individual brands (Simonin

& Ruth, 1998) and elicit more positive perceptions than single-brand extensions (Park et al., 1996), little is known about how investors react to introductions of co-branded products. A majority of published research studies use an experimental approach to measure consumers' perceptions and awareness of co-branded products and constituent brands. There is a dearth of research on the stock market impact of co-branding. Positive evaluations obtained in lab settings may not necessarily translate into actual profits in an intensely competitive environment. The financial gains of co-branding in the stock market could help us pinpoint the value of co-branding activities more accurately. Moreover, this study investigates the effects of certain co-branding characteristics on financial returns in the short-term and the long term.

The rest of the paper is structured as follows. Section II provides the background and definition of co-branding activities. A classification of co-branding is presented because there are other brand alliance terms in the literature to describe the cooperation between brands. This section also provides a review of the literature and summarizes the findings with respect to the links between co-branding and consumer attitudes and the links between branding activities that include co-branding and firms' financial performance. Following this, section III outlines the conceptual foundation of the theoretical framework and proposes hypotheses about the stock market's reaction to co-branding announcements. Finally, section IV provides the conclusion.

## 2- Definitions and Literature Review

### 2-1- Definitions

Forming an alliance with other established brands has become a widely used business strategy since the 1980s, when brand equity became an essential measure of businesses' real value. Brand alliances can take many forms, from product bundling to dual branding to co-branding.

First, product bundling is a strategy in which two or more products are sold together for one price ((Gaeth et al., 1991); (Yadav, 1994)). The bundle can be either composed of physical goods like a laptop with its peripherals or intangible services such as a holiday package including an airline ticket and a city tour guide. While in many instances, the components of the bundle carry the same brand, there are cases in which different brands are sold together in one package (e.g., fragrance or skincare multi-brand packs sold by Sephora). Product bundling is also encountered in promotions, where typically one branded product is offered for free with the purchase of another branded product (Varadarajan, 1986)(Tahami et al., 2016).

Second, "Advertising alliance" (or co-advertising, joint advertising) is characterized by the use of two brands on a promotional campaign delivered by an advertisement (Samu et al., 1999). The joint advertising of Axe anti-sweat spray and Coca-Cola zero is a typical case. This anti-sweat spray brand signals its functional performance by borrowing the associations of the "cool drink" of Coca-Cola. Other example include the Kellogg cereals with Tropicana fruit juice (Samu et al., 1999) .

Third, "Joint-sales promotion" represents the cooperation of the promotional resource shared by two or more brands designed to seek an opportunity for sales growth. In this specific type of alliance, each of the brands (products) can be either adopted independently or promoted together for complimentary use. The examples include the Campbell soup with the Nabisco saltine crackers (Varadarajan, 1986), the Bacardi Rum with Coca-Cola, and a menu including a handmade pizza with a Coca-Cola.(Tahami et al., 2019)

Fourth, "dual branding" concerns an arrangement in which two brands share the same location and consumers, therefore, can purchase their products under the same roof (e.g., Kentucky Fried Chicken-A&W restaurants (Levin et al., 1996). Fifth, "brand extension" refers to using an existing brand name to launch a new product (Aaker & Keller, 1990). (Kotler & Keller, 2006) claimed that there exist two types of brand extension strategies, namely the "line extension" and "category extension." Line extension incorporates the established brand name into the firm's existing product category (e.g., Coca-Cola introduced the vanilla flavor). The extended product differs from the original brand in several minor characteristics such as flavors and sizes (Kapferer, 1998)(Tahami & Fakhravar, 2020c). On the contrary, the "category extension" uses the current brand name to access a different product category (e.g., the Apple iPod).

Several researchers ((Park et al., 1996); (Hadjicharalambous, 2006); (Helmig et al., 2008)(Fakhravar, n.d.; Tahami & Fakhravar, 2020a, 2020b, 2020b)) argued that co-branding and brand extension are similar in their purposes: both of them are utilized to reduce the risk of failures of new products by capitalizing on the existing equities of the parent brand(s) and by transferring the existing brand associations to the new product. However, (Leuthesser et al., 2003) claimed that co-branding is sometimes a more effective strategy than brand extension because co-branding has a smaller possibility to dilute the attitudes toward the partnering brands and to damage the allying brands' images. (Hadjicharalambous, 2006) and (Helmig et al., 2008) claimed that the only difference between the two strategies is the number of constituent brands involved: brand extension is characterized by the use of a single brand while co-branding involves a combination of two brands.(Fakhravar, 2020a, 2021)

Table 1 summarizes the distinctions between co-branding activities and other types of branding strategies listed above.

*Table 1 - distinctions between co-branding activities and other types of branding strategies*

| Strategy | Characteristics | Example | Difference from co-branding |
|---|---|---|---|
| **Product bundling** | Combined offer of two or more goods in a package with one total price | Skincare multi-brand packs sold by Sephora | No simultaneous branding of a single physical product by two brands |
| **Advertising alliance** | Simultaneous mention of different suppliers of different products in one advertisement | Nike sport shoes and Apple iPod player | |
| **Joint sales promotions** | Timely, limited appearance of two independent brands in Promotional activities | Reebok® & Pepsi® | |
| **Dual branding** | Common usage of a store location (shop in shop concept) | Burger King® & Shell™ | |
| **Brand extension** | Extension of a brand to a new product in either a new or an existing product category | Boss brand transfer from clothes to perfumes | Equals co-branding, if the new product is branded by two brands simultaneously |

## 2-2 Literature Review

Prior researchers have measured the success of co-branding mostly under two main frameworks, namely the strategic alliance framework and the consumer behavior framework. Due to the abundant literature in analyzing the successful strategic alliances (e.g.,(Angle & Perry, 1981); (Devlin & Bleackley, 1988); (Mohr &

Spekman, 1994)(Ouabira & Fakhravar, 2021)), some marketing researchers have utilized the framework of strategic alliance to examine the success of co-branding. In their studies, the term "alliance success" is often referred to a "successful formation of a co-branding alliance". Under this framework, most research explores the strategic intents of one brand partner to form a co-branding alliance from two research streams (Fakhravar, 2022). One stream of work emphasizes the signals of a brand (names) (e.g., (Rao & Ruekert, 1994); (Washburn et al., 2000); (Bengtsson & Servais, 2005)) while the other stream of literature focuses on the organizational characteristics and the mutual benefits.(Natarajan et al., 2021)

Signaling theory (Spence, 1973) has been adopted by several scholars (e.g., (Rao & Ruekert, 1994); (Washburn et al., 2000); (Bengtsson & Servais, 2005)) to explain the function of the brand name in a co-branding alliance. (Rao & Ruekert, 1994) is a seminal piece in applying the signaling theory into co-branding research. They treated co-branding as a strategic alliance and analyzed why the partnership is formed by combining their brand names (i.e., as compared to the cooperation in only the R&D level). They argued that the name of each brand could serve as a signal for product quality to the consumers(Series, n.d.). Hence each brand can offer its brand name as a benefit to the other. In another influential paper, (Washburn et al., 2000) stated that consumers could transfer the existing brand associations of the allying brands to the alliance, and the strength of the association can represent the magnitude of brand equity. In summary, the signaling theory is employed to investigate how the brand name can serve as a benefit to be offered or a single piece of information to be displayed in a partnership.

However, using the framework of a strategic alliance to study co-branding has an intrinsic limitation: the behavior of consumers is not fully taken into account (Hadjicharalambous, 2006). Since co-branding is a combination of two brands and ultimately, the brands are "owned" in the minds and hearts of consumers (Leuthesser et al., 2003), consumer evaluation is an indispensable issue in co-branding research. That is to say, co-branding should also be studied from the consumer behavior perspective.(Tahami & Fakhravar, 2020b)

In contrast to a small number of studies from the strategic alliance framework, a rich vein of literature has investigated this topic from the consumer behavior framework (Leuthesser et al., 2003). With a large base of attitude research as their background, the behavioral researchers often use "attitudinal acceptance" to measure the effectiveness of a co-branding strategy. In particular, a great deal of attention has been paid to consumers' attitudes toward the co-brand (e.g., (Park et al., 1996); (Samu et al., 1999); (Desai & Keller, 2002); (James et al., 2006)) and the post-exposure attitudes toward each of the allying brands (e.g., (Park et al., 1996); (Rodrigue & Biswas, 2004); (Abbo, 2006)).

"Fit" is the most important factor in determining the attitudinal acceptance of the co-brand. (Aaker & Keller, 1990) measured the concept of "fit" in three dimensions, namely complement, substitute, and transfer. Complement refers to that a particular need that can be satisfied by a joint consumption of two products in different categories. Substitute pertains to the replacement of one product by the other to fulfill the same desire. The above two dimensions are discussed from a demand-side perspective but, from the supply-side of view, transfer refers to the "perceived ability of any firm operating in the first product class to make a product in the second product class" (Aaker & Keller, 1990).

(Park et al., 1996) was the first researcher to use "fit" in co-branding research. They stated that the attribute complementarity (i.e., product-fit) plays an important role, and they defined the term "complementarity" between the allying brands (e.g., say brand A and B) as:

1- Both brands have a set of relevant attributes (e.g., good taste and low-calories).

2- Two brands differ in attribute salience such that the attribute, which is not salient to one, is salient to the other (e.g., brand A is salient in good taste and brand B is salient in low-calories).

3- The brand for which the attribute is salient has a higher perceived performance level on that attribute (e.g., brand A has a higher level on "good-taste" and brand B has a higher performance level on "low-calories").

In addition to the product fit between the allying brands, the brand fit (i.e., consistent brand images) is also proved to have a considerable positive influence on the attitudes toward the co-brand (Simonin & Ruth, 1998). (Simonin & Ruth, 1998) further found that the degree of brand familiarity moderates each brand's contribution to the evaluation of co-brand. Other than the concept of "fit", several other factors such as country-of-origin effect (Voss & Tansuhaj, 1999), association transfer (James et al., 2006), product involvements and brand orientations (Huber, 2005), and the strength of reciprocal effects (Rodrigue & Biswas, 2004) can also affect consumer attitudes toward the co-brand.

Although co-branding has been extensively researched in the marketing field, positive evaluations obtained in experimental settings do not necessarily suggest actual profits in an intensely competitive marketplace. One manner in which the true value of co-branding in the marketplace can be established is by examining the stock market's reaction to the introduction of co-branded products.

### 2-2-1 Branding and Stock Market Returns

Brands are viewed as intangible assets that generate future cash flows (Aaker & Jacobson, 1994) or reduce the volatility of future cash flows (Ambler, 2003). As brand equity is a complex concept, (Keller & Lehmann, 2006) suggested that branding-shareholder value is reflected in three perspectives: customer-based equity, product-market brand equity, and financial-based brand equity. All three components of brand equity have been found to be able to drive firm value (e.g., (Madden et al., 2006)). Many authors argue that financial-based brand equity as a metric of brand equity goes well beyond short-term sales, profits, and market share, and it is growing in appeal. (Simon & Sullivan, 1993) estimated a firm's brand equity that is based on the financial market value of the firm. They defined brand equity as the incremental cash flows which accrue to branded products over unbranded products. Their approach provided an objective value of a firm's brands that is related to the determinants of brand equity.

(Barth et al., 1998) used simultaneous equations estimation to investigate the relationships between brand value and returns and accounting variables. They showed that brand value estimates are positively associated with advertising expense, operating margin, and market share. Their findings suggested brand value estimates provide significant explanatory power for prices incremental to these variables and to recognized brand assets and analysts' earnings forecasts.

(Madden et al., 2006) investigated the link between shareholder value and brand assets and provided evidence pertaining to how marketing affects firm performance. Using the Fama-French method, the authors showed that, when market share and firm size are considered, strong brands not only deliver greater returns to stockholders but also help reduce the risk. Their findings provided a more comprehensive perspective by supporting the importance of marketing function as the processes that create firm value.

As investors view incremental information on branding activities as contributing to estimating future cash flows, the challenges emerge for marketers and researchers to assess and communicate the value created by corporate branding strategies on shareholder value. Research supports that firm value is linked to corporate naming strategies that influence brand awareness can change brand equity. Corporate activities related to social responsibility are also suggested to have an impact on firm value through building brand images. Moreover, many studies have documented the manner in which mergers and acquisitions (M&A) and brand portfolio strategies impact firm value. (Srivastava et al., 1998) developed a conceptual framework of the marketing-finance interface and propose that marketing tasks involve developing and managing market-based assets that include customer relationships, channel relationships, and partner relationships. Market-based assets such as brands can increase shareholder value through accelerating cash flows, lowering the volatility and vulnerability of cash flows, and enhancing the residual value of cash flows.

It is possible that brands can benefit or suffer from extensions. For instance, brand extensions may restrict the financial value creation of the firm because brand extensions may preclude opportunities that are provided only through new and unconnected brand offerings (Aaker & Keller, 1990). However, there is a dearth of research on the link between brand extensions and changes in financial returns. (Lane & Jacobson, 1995) used the event study method to investigate the financial returns of brand extension announcements and found that stock market response depends on brand equity components, including brand familiarity and attitude towards the extension brands. Their analysis indicated that brand equity characteristics significantly influence the success of brand extensions.

## 2-2-2 Literature Gap

While a large body of research focuses on consumers' assessment of cobranding ((Rao & Ruekert, 1994); (Rao et al., 1999); (Voss & Gammoh, 2004)), "the lack of firm-side research is a fundamental limitation in the extant brand alliance literature" (Gammoh & Voss, 2011). Consistent with this view, our literature review in Table 2 suggests a gap in knowledge about the value effect of co-branding strategy. (Cao & Sorescu, 2013) is the only study that examines the effect of cobranded new product introductions on firm stock performance. They find that the stock market rewards the manufacturer of cobranded products with a 1% increase in stock returns (Nguyen et al., 2020). However, this research stops short of uncovering how cobranding creates value for both partners. Several important questions remain unanswered: 1) Does cobranding create value for all partnering firms? 2) What factors drive the value creation and distribution of co-branding?

*Table 2. Summary of literature on the effect of brand alliances on stock market performance*

| Article | Research context | Theoretical lenses | Performance metric | Key findings |
|---|---|---|---|---|
| **Das et al. (1988)** | 70 marketing and 49 technological alliances in 18 industries | Transaction cost economics | CAR | Technological alliances create greater CARs for partnering firms than marketing alliances. CARs are negatively correlated with firm profitability and size. |
| **Luo et al. (2007)** | Surveys of 387 executives in 8 industries | Resource capability | Return on equity (ROE) | Intensity of a firm's alliances with its competitors has a curvilinear effect on ROE. |
| **Kalaignanam et al. (2007)** | 167 new product development alliances in information technology and telecommunication industries | Power in interfirm relationship | CAR | Partnering firms obtain positive CARs, but the gain is asymmetric between larger and smaller firms. While a broad scope alliance enhances financial gains for larger firms, a scale R&D alliance contributes to the financial gains for smaller firms. |
| **Swaminamthan and Morrman (2009)** | 230 marketing alliances in computer software industry | Signaling | CAR | Marketing alliances create positive CARs for partnering firms. |
| **Cao and Sorescu (2013)** | 316 announcements of cobranded new products in the CPG industry | Signaling | CAR | Average CARs of the manufacturing firm to the announcement of cobranded new products is approximately 1.0%. |
| **Borah and Tellis (2014)** | 3,522 announcements of make, buy, or ally for innovations | Organizational learning | CAR | Announcements to make or ally for innovation generate positive and higher CARs than announcements to buy, which generate negative CARs. |

## 3- Hypothesis Development

The conceptual framework is depicted in Figure 1, and the concepts and hypotheses were elaborated on next.

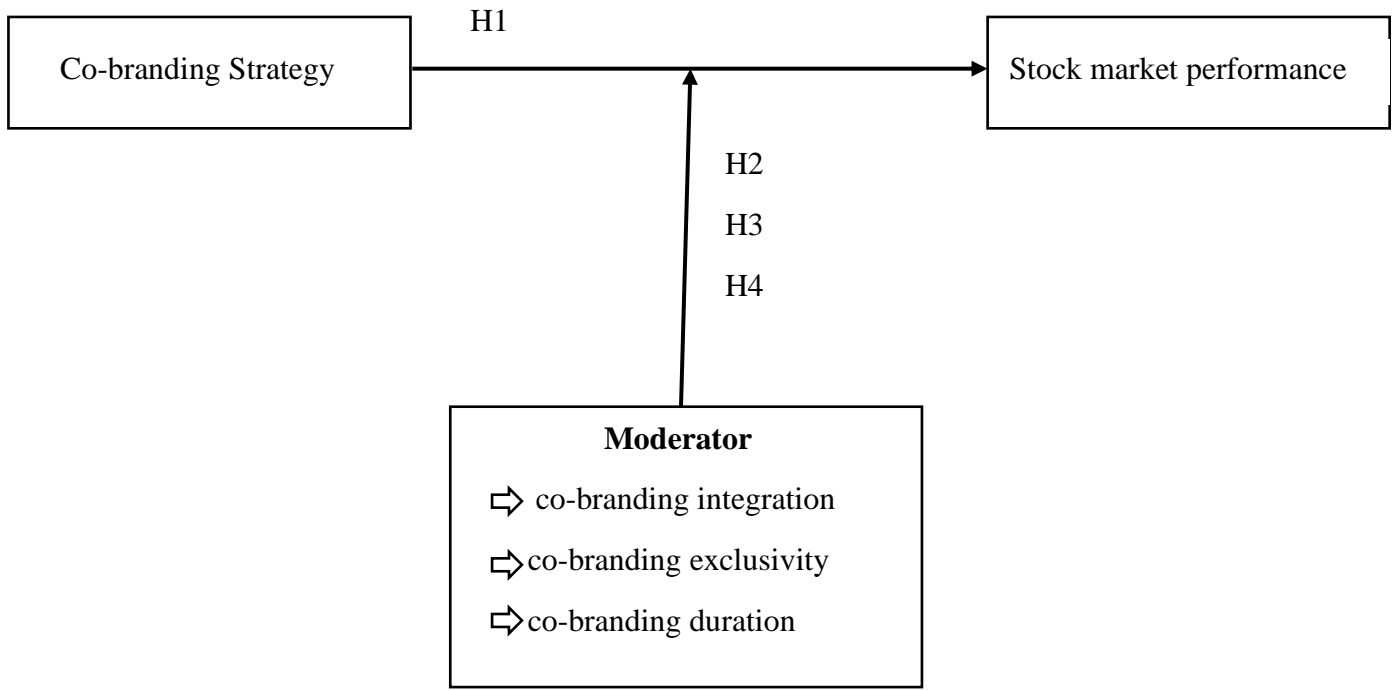

*Figure 1. Conceptual Framework*

## 3-1 Impact of co-branding on stock performance

New co-branded products share two unique features that are likely to be viewed more favorably by investors when compared to single-branded products. First, co-branding can signal quality to consumers (Rao et al., 1999) and can improve consumers' attitudes toward individual partner brands, with positive brand association spillovers documented both from the individual brands to the co-branded product and vice versa (Simonin & Ruth, 1998). With greater credibility of product quality, co-branded products can also command a premium price (Venkatesh & Mahajan, 1997) and elicit more positive perceptions than single-brand extensions ((Desai & Keller, 2002); (Park et al., 1996)). Collectively, these findings suggest that co-branded products are likely to be viewed more favorably by consumers and generate higher cash flows.

Co-branding alliances also offer the partner firms an opportunity to improve operational efficiencies. Co-branding partners can gain access to new markets and share each other's resources in terms of manufacturing, managerial knowledge, and advertising. Prior research shows that the market reaction is positive in the case of a new product or technological alliances (e.g., (Kalaignanam et al., 2007)), suggesting that investors anticipate higher future cash flows as a result of technology transfers. Because co-branding alliances involve the creation and commercialization of a new product, they draw upon characteristics of both marketing and technological alliances. Consequently, such alliances should not only generate higher future cash flows (similar to those generated by new product alliances) but also reduce the uncertainty associated with these cash flows by leveraging the equities of the two partner brands. Therefore, it is expected that the stock market reaction to the announcement of co-branded products will be positive.

*H1: Co-branding event announcements generate positive abnormal returns in the stock market.*

### 3-1 Moderating effect of co-branding attributes

Different attributes of co-branding structure could also impact an involved firm's abnormal returns. According to (Newmeyer et al., 2014), there are three co-branding structures, namely co-branding integration, co-branding exclusivity, and co-branding duration. Integration refers to the extent to which the partnering brands are intertwined in form and function. In such a relationship, the brands can be presented together in a multitude of forms, with the degree of integration ranging from very low where the brands are almost entirely self-standing and separate in form and function to very high, with the brands completely fused together such that it is practically impossible to separate them in form and function. In high integration, multiple brands are paired together to make a complete product, and the highest level of utility is achieved when the brands are used jointly. In contrast, low integration is the joint presentation of brands that largely remain separate in form, and joint use may be desirable but certainly not necessary. Exclusivity is defined as the number of partners with whom the focal brand pursues a co-branding arrangement. Greater exclusivity is when the focal brand has a single (or few) partner brand(s). For example, the Apple iPhone initially had an exclusive alliance with AT&T. Less exclusive co-branding is one in which the focal brand has many partners (Intel's co-branding with many PC manufacturers). In an exclusive arrangement, all of the focal brand's performance information is derived from the single partnership or the individual performance of the parent brand. In a less exclusive co-branding arrangement, consumers have broader focal brand performance information obtained from multiple and possibly diverse partnerships ((Das & Teng, 1998); (Mowery et al., 1996)).

Duration is defined as the length of time for which the cobranding arrangement lasts (e.g.,(Das & Teng, 1998); (Provan & Gassenheimer, 1994)). The relationship could be short-term (Disney's Lion King co-promotion with Burger King for that movie only) or long-term, encompassing multiple generations (Pixar's partnership with Disney to co-produce multiple movies, finally leading to the merger of Pixar and Disney).

According to the existing research, high co-branding integration, long co-branding duration, and high co-branding exclusivity are found to generate a greater impact on brand evaluation. Thus, the following three hypotheses are proposed.

*H2: Co-branding event announcements generate greater positive abnormal returns in the stock market when the partnering brands are highly integrated.*

*H3: Co-branding event announcements generate greater positive abnormal returns in the stock market when the co-branding structure is more exclusive.*

*H4: Co-branding event announcements generate greater positive abnormal returns in the stock market when co-branding duration is long term.*

## 4- Conclusion

In this paper, we investigated the relation between co-branding event announcement and abnormal returns in the stock market for the partnering brands. We argued that co-branding event announcement generates positive abnormal returns in the stock market. Furthermore, we investigated the moderating impact of co-branding structure attributes on the relation between co-branding event announcement and abnormal returns. We claimed that higher co-branding integration, greater co-branding exclusivity, and longer co-branding duration generate a greater positive abnormal return for the partnering firms.

# Reference

Aaker, D. A., & Jacobson, R. (1994). The Financial Information Content of Perceived Quality. *Journal of Marketing Research*, *31*(2), 191–201. https://doi.org/10.2307/3152193

Aaker, D. A., & Keller, K. L. (1990). Consumer Evaluations of Brand Extensions. *Journal of Marketing*, *54*(1), 27–41. https://doi.org/10.1177/002224299005400102

Abbo, M.-H. (2006). *An Exploratory Study on the Impact of Two Ingredient Branding Strategies on the Host Brand*.

Angle, H. L., & Perry, J. L. (1981). An Empirical Assessment of Organizational Commitment and Organizational Effectiveness. *Administrative Science Quarterly*, *26*(1), 1–14. https://doi.org/10.2307/2392596

Barth, M. E., Clement, M. B., Foster, G., & Kasznik, R. (1998). Brand Values and Capital Market Valuation. *Review of Accounting Studies*, *3*(1), 41–68. https://doi.org/10.1023/A:1009620132177

Bengtsson, A., & Servais, P. (2005). Co-branding on industrial markets. *Industrial Marketing Management*, *34*, 706–713.

Besharat, A., & Langan, R. (2014). Towards the formation of consensus in the domain of co-branding: Current findings and future priorities. *Journal of Brand Management*, *21*(2), 112–132. https://doi.org/10.1057/bm.2013.25

Cao, Z., & Sorescu, A. (2013). Wedded Bliss or Tainted Love? Stock Market Reactions to the Introduction of Cobranded Products. *Marketing Science*, *32*(6), 939–959. https://doi.org/10.1287/mksc.2013.0806

Das, T. K., & Teng, B.-S. (1998). Resource and risk management in the strategic alliance making process. *Journal of Management*, *24*(1), 21–42. https://doi.org/https://doi.org/10.1016/S0149-2063(99)80052-X

Desai, K. K., & Keller, K. L. (2002). The Effects of Ingredient Branding Strategies on Host Brand Extendibility. *Journal of Marketing*, *66*(1), 73–93. https://doi.org/10.1509/jmkg.66.1.73.18450

Devlin, G., & Bleackley, M. (1988). Strategic alliances—Guidelines for success. *Long Range Planning*, *21*(5), 18–23. https://doi.org/https://doi.org/10.1016/0024-6301(88)90101-X

Fakhravar, H. (n.d.). *Fuzzy inventory model with receiving reparative order and considering imperfect quality items*.

Fakhravar, H. (2020a). *Application of Failure Modes and Effects Analysis in the Engineering Design*.

Fakhravar, H. (2020b). Quantifying Uncertainty in Risk Assessment using Fuzzy Theory. *ArXiv E-Prints*, arXiv:2009.09334.

Fakhravar, H. (2021). Application of Failure Modes and Effects Analysis in the Engineering Design Process. *ArXiv Preprint ArXiv:2101.05444*.

Fakhravar, H. (2022). Combining heuristics and Exact Algorithms: A Review. *ArXiv Preprint ArXiv:2202.02799*.

Gaeth, G. J., Levin, I. P., Chakraborty, G., & Levin, A. M. (1991). Consumer evaluation of multi-product bundles: An information integration analysis. *Marketing Letters*, *2*(1), 47–57. https://doi.org/10.1007/BF00435195

Helmig, B., Huber, J.-A., & Leeflang, P. S. H. (2008). Co-branding: The State of the Art. *Schmalenbach Business Review*, *60*(4), 359–377. https://doi.org/10.1007/BF03396775

James, D. O., Lyman, M., & Foreman, S. K. (2006). Does the tail wag the dog? Brand personality in brand alliance evaluation. *Journal of Product & Brand Management*, *15*(3), 173–183. https://doi.org/10.1108/10610420610668612

Kalaignanam, K., Shankar, V., & Varadarajan, R. (2007). Asymmetric New Product Development Alliances: Win-Win or Win-Lose Partnerships? *Management Science*, *53*(3), 357–374. https://doi.org/10.1287/mnsc.1060.0642

Keller, K. L., & Lehmann, D. (2006). Brands and Branding: Research Findings and Future Priorities. *Marketing Science*, *25*, 740–759.

Keränen, J., Piirainen, K. A., & Salminen, R. T. (2012). Systematic review on B2B branding: research issues and avenues for future research. *Journal of Product & Brand Management*, *21*(6), 404–417. https://doi.org/10.1108/10610421211264892

Lane, V., & Jacobson, R. (1995). Stock Market Reactions to Brand Extension Announcements: The Effects of Brand Attitude and Familiarity. *Journal of Marketing*, *59*(1), 63–77. https://doi.org/10.2307/1252015

Leuthesser, L., Kohli, C., & Suri, R. (2003). 2+2=5? A framework for using co-branding to leverage a brand. *Journal of Brand Management*, *11*(1), 35–47. https://doi.org/10.1057/palgrave.bm.2540146

Levin, A. M., Davis, J. C., & Levin, I. (1996). Theoretical and Empirical Linkages

Between Consumers' Responses to Different Branding Strategies. *ACR North American Advances*.

Madden, T. J., Fehle, F., & Fournier, S. (2006). Brands matter: An empirical demonstration of the creation of shareholder value through branding. *Journal of the Academy of Marketing Science*, *34*(2), 224. https://doi.org/10.1177/0092070305283356

Mohr, J., & Spekman, R. (1994). Characteristics of partnership success: Partnership attributes, communication behavior, and conflict resolution techniques. *Strategic Management Journal*, *15*(2), 135–152. https://doi.org/https://doi.org/10.1002/smj.4250150205

Mowery, D. C., Oxley, J. E., & Silverman, B. S. (1996). Strategic alliances and interfirm knowledge transfer. *Strategic Management Journal*, *17*(S2), 77–91. https://doi.org/https://doi.org/10.1002/smj.4250171108

Natarajan, G., Ng, E. H., & Katina, P. F. (2021). *SYSTEMS STATISTICAL ENGINEERING–HIERARCHICAL FUZZY CONSTRAINT PROPAGATION*.

Newmeyer, C. E., Venkatesh, R., & Chatterjee, R. (2014). Cobranding arrangements and partner selection: a conceptual framework and managerial guidelines. *Journal of the Academy of Marketing Science*, *42*(2), 103–118. https://doi.org/10.1007/s11747-013-0343-8

Nguyen, H. T., Ross, W. T., Pancras, J., & Phan, H. V. (2020). Market-based drivers of cobranding success. *Journal of Business Research*, *115*, 122–138. https://doi.org/https://doi.org/10.1016/j.jbusres.2020.04.046

Ouabira, M. M., & Fakhravar, H. (2021). Effective Project Management and the Role of Quality Assurance throughout the Project Life Cycle. *European Journal of Engineering and Technology Research*, *6*(5).

Park, C. W., Jun, S. Y., & Shocker, A. D. (1996). Composite Branding Alliances: An Investigation of Extension and Feedback Effects. *Journal of Marketing Research*, *33*(4), 453–466. https://doi.org/10.1177/002224379603300407

Paydas Turan, C. (2021). Success drivers of co-branding: A meta-analysis. *International Journal of Consumer Studies*, *n/a*(n/a). https://doi.org/https://doi.org/10.1111/ijcs.12682

Provan, K. G., & Gassenheimer, J. B. (1994). SUPPLIER COMMITMENT IN RELATIONAL CONTRACT EXCHANGES WITH BUYERS: A STUDY OF INTERORGANIZATIONAL DEPENDENCE AND EXERCISED POWER*. *Journal of Management Studies*, *31*(1), 55–68.

https://doi.org/https://doi.org/10.1111/j.1467-6486.1994.tb00332.x

Rao, A. R. (1997). Strategic brand alliances. *Journal of Brand Management*, *5*(2), 111–119. https://doi.org/10.1057/bm.1997.37

Rao, A. R., Qu, L., & Ruekert, R. W. (1999). Signaling Unobservable Product Quality through a Brand Ally. *Journal of Marketing Research*, *36*(2), 258–268. https://doi.org/10.2307/3152097

Rodrigue, C. S., & Biswas, A. (2004). Brand alliance dependency and exclusivity: an empirical investigation. *Journal of Product & Brand Management*, *13*(7), 477–487. https://doi.org/10.1108/10610420410568417

Samu, S., Krishnan, H. S., & Smith, R. E. (1999). Using Advertising Alliances for New Product Introduction: Interactions between Product Complementarity and Promotional Strategies. *Journal of Marketing*, *63*(1), 57–74. https://doi.org/10.1177/002224299906300105

Series, O. D. L. (n.d.). *illuminator*.

Simon, C. J., & Sullivan, M. W. (1993). The Measurement and Determinants of Brand Equity: A Financial Approach. *Marketing Science*, *12*(1), 28–52.

Simonin, B. L., & Ruth, J. A. (1998). Is a Company Known by the Company It Keeps? Assessing the Spillover Effects of Brand Alliances on Consumer Brand Attitudes. *Journal of Marketing Research*, *35*(1), 30–42. https://doi.org/10.2307/3151928

Spence, M. (1973). Job Market Signaling. *The Quarterly Journal of Economics*, *87*(3), 355–374. https://doi.org/10.2307/1882010

Srivastava, R. K., Shervani, T. A., & Fahey, L. (1998). Market-Based Assets and Shareholder Value: A Framework for Analysis. *Journal of Marketing*, *62*(1), 2–18. https://doi.org/10.2307/1251799

Tahami, H., & Fakhravar, H. (2020a). A Fuzzy Inventory Model Considering Imperfect Quality Items with Receiving Reparative Batch and Order. *European Journal of Engineering and Technology Research*, *5*(10), 1179–1185.

Tahami, H., & Fakhravar, H. (2020b). A fuzzy inventory model considering imperfect quality items with receiving reparative batch and order overlapping. *ArXiv Preprint ArXiv:2009.05881*.

Tahami, H., & Fakhravar, H. (2020c). Multilevel Reorder Strategy-based Supply Chain Model. *5th North American Conference on Industrial Engineering and*

*Operations Management (IEOM), Michigan, USA*.

Tahami, H., Mirzazadeh, A., Arshadi-khamseh, A., & Gholami-Qadikolaei, A. (2016). A periodic review integrated inventory model for buyer's unidentified protection interval demand distribution. *Cogent Engineering*, *3*(1), 1206689.

Tahami, H., Mirzazadeh, A., & Gholami-Qadikolaei, A. (2019). Simultaneous control on lead time elements and ordering cost for an inflationary inventory-production model with mixture of normal distributions LTD under finite capacity. *RAIRO-Operations Research*, *53*(4), 1357–1384.

Varadarajan, P. "Rajan." (1986). Horizontal Cooperative Sales Promotion: A Framework for Classification and Additional Perspectives. *Journal of Marketing*, *50*(2), 61–73. https://doi.org/10.2307/1251600

Venkatesh, R., & Mahajan, V. (1997). Products with Branded Components: An Approach for Premium Pricing and Partner Selection. *Marketing Science*, *16*(2), 146–165.

Voss, K. E., & Gammoh, B. S. (2004). Building Brands through Brand Alliances: Does a Second Ally Help? *Marketing Letters*, *15*(2), 147–159. https://doi.org/10.1023/B:MARK.0000047390.01552.a2

Voss, K. E., & Tansuhaj, P. (1999). A Consumer Perspective on Foreign Market Entry. *Journal of International Consumer Marketing*, *11*(2), 39–58. https://doi.org/10.1300/J046v11n02_03

Washburn, J. H., Till, B. D., & Priluck, R. (2000). Co-branding: brand equity and trial effects. *Journal of Consumer Marketing*, *17*(7), 591–604. https://doi.org/10.1108/07363760010357796

Yadav, M. S. (1994). How Buyers Evaluate Product Bundles: A Model of Anchoring and Adjustment. *Journal of Consumer Research*, *21*(2), 342–353.

Yahoodik, S., Tahami, H., Unverricht, J., Yamani, Y., Handley, H., & Thompson, D. (2020). Blink Rate as a Measure of Driver Workload during Simulated Driving. *Proceedings of the Human Factors and Ergonomics Society Annual Meeting*, *64*(1), 1825–1828.

Yu, Y., Rothenberg, L., & Moore, M. (2021). Exploring young consumer's decision-making for luxury co-branding combinations. *International Journal of Retail & Distribution Management*, *49*(3), 341–358. https://doi.org/10.1108/IJRDM-12-2019-0399